\documentclass[
reprint,
superscriptaddress,
groupedaddress,
showpacs,
showkeys,
 amsmath,amssymb,
 aps,
prb,
floatfix,
]{revtex4-2}

\usepackage{graphicx}
\usepackage{dcolumn}
\usepackage{bm}

\usepackage{amsmath}
\usepackage{graphicx}
\usepackage{booktabs}
\usepackage{siunitx}
\usepackage{graphicx}
\usepackage{dcolumn}
\usepackage{bm}
\usepackage{multirow}
\usepackage{placeins}

\begin{document}


\title{Charge Ordering in out-of-plane Boron Doped Reduced Graphene Oxide}
\author{Saikat Sarkar}
\affiliation{Thin Film and Nanoscience Laboratory, Department of Physics, Jadavpur University, Kolkata-700032, India.}
\author{Rajarshi Roy}
\affiliation{Department of Semiconductors, Institute of Physics (FZU), Cukrovarnická 112/10, 16200 Praha 6}
\author{Bikram Kumar Das}
\affiliation{Thin Film and Nanoscience Laboratory, Department of Physics, Jadavpur University, Kolkata-700032, India.}
\affiliation{Present address: Modelling and Simulation in Life and Materials science, BCAM - Basque Center for Applied Mathematics, Alameda de Mazarredo 14, E-48009, Bilbao, Spain}
\author{Suman Chatterjee}
\affiliation{Quantum Electronics Lab, Dept. of ECE, IISc Bangalore, Bangalore-560012, India.}
\author{Kalyan Kumar Chattopadhyay}\email{kalyan$_$chattopadhyay@yahoo.com}
\affiliation{Department of Physics, Jadavpur University, Kolkata-700032, India.}
\affiliation{School of Materials Science and Nanotechnology, Jadavpur University, Kolkata-700032, India.} 
 \date{\today}

\begin{abstract}
Symmetry-breaking phase transitions analogous to superconductivity (SC), charge ordering (CO) etc. in metal-intercalated graphene are favorable resulting from modified electronic and phonon band structures. Strong carrier-lattice interaction evolved from the out-of-plane soft vibrations with accumulation of charges at the out-of-plane region, can set a favorable environment for CO in graphene system. Here, we employ boron-doped reduced graphene oxide (BG) to acquire charge-ordered state above a transition temperature, $T_1 \approx 97.5K$. Signatures of this state are identified using ab-initio simulations and low temperature electrical transport measurements. The out-of-plane boron groups play a crucial role in reinforcing the electron-phonon coupling (EPC) allowing an ordered-state transition. Temperature-dependent Raman spectroscopy further supports the emergence of ordering. Key characterization techniques (X-ray diffraction, Raman spectra etc.) are used to quantify the EPC interaction and associated factors like tensile strain, boundary defects, etc. affecting charge ordering with doping. Additionally, we find interesting electric field dependency on the CO in this non-metallic, light-atom-doped chemically derived graphene. 
\end{abstract}

\pacs{Valid PACS appear here}
\maketitle

\section{\label{sec:level1}Introduction}

Recent advancement of doped graphene systems towards obtaining a strongly correlated electron system unfolds a promising field for graphene as well as for other two-dimensional materials (2D) \cite {Ehlen2020,Wang2022,Long2016,Gao2021,Zhang2016,Chatterjee2015,Gao2020,Baraghani2022,Tsen2015}. Since the last decade, intercalation of alkali metals \cite{Ehlen2020,Xue2012,Ludbrook2015}, and other alkaline earth elements \cite{Ichinokura2016,Chapman2016} in graphene has become an area of interest for inducing macroscopic quantum phenomena like SC \cite{Xue2012,Ludbrook2015,Ichinokura2016,Chapman2016}, CO \cite{Shimizu2015,Rahnejat2011}. However, the origin of both superconductivity and CO in 2D materials largely remain ambiguous as several microscopic factors such as Fermi surface nesting (FSN) \cite{Inosov2008,Borisenko2008,Borisenko2009}, electron-phonon coupling (EPC) \cite{Johannes2008,Varma1983,Sanna2022}, formation of exciton pair \cite{Kidd2002,Kogar2017,Jrome1967} and  exciton-phonon interaction \cite{vanWezel2010}either separately or collectively assist together to construct such correlated phases. In fact, the electron-electron interactions weigh heavily as a predominant term in the CO process \cite{Watanabe2015,Jang2019,Horovitz1985}. Coexistence of these phases (SC, CO) in layered 2D materials, \cite{Valla2004,Weber2011,Lian2019,Ryu2018},is found to act either in or out of favor with each other. In such cases, it is found that this EPC interaction plays a pivotal role to develop an attractive term to surmount the Coulombic part between the charge carriers in layered materials. 
In general, the description of the EPC is given by an equation as,
\begin{equation}
\lambda = \frac{{N}\left(E_{F}\right)D^{2}}{M{\omega_{ph}^2}} 
\end{equation}                                              
,where $N(E_F)$ is the electronic density of states (DOS) per spin at Fermi level $(E_F)$, D is the deformation potential, M is the effective atomic mass, $\omega_{ph}$ is the phonon frequency \cite{Profeta2012}. Thus, in the context of pure graphene due to the vanishing density of states at the Dirac point there is non-existent EPC interaction. Nonetheless, one can induce and strengthen this coupling by manipulating these contributing factors in equation (1) in doped system. Electron or hole doping of graphene directly increases the DOS near Fermi level by shifting the Fermi level to the conduction band or valance band respectively. But doping in the plane of lattice of the graphene, i.e., in-plane doping cannot elevate the value of the EPC, as extremely energetic in-plane phonons amplify the $\omega_{ph}$ term in the denominator of equation (1). However, intercalation of graphene can contribute to reinforce the EPC in two ways. At first by increasing the DOS at $(E_F)$ and then adding a considerable amount to the out-of-plane vibrations associated with the softer phonon modes. So, similar like metal-intercalated graphene, doping at the interlayer (IL) region of layered graphene would be the appropriate technique to develop CO phase by enhancing the EPC \cite{Rahnejat2011,Shimizu2015}. 

\begin{figure*}
\includegraphics[width=\textwidth]{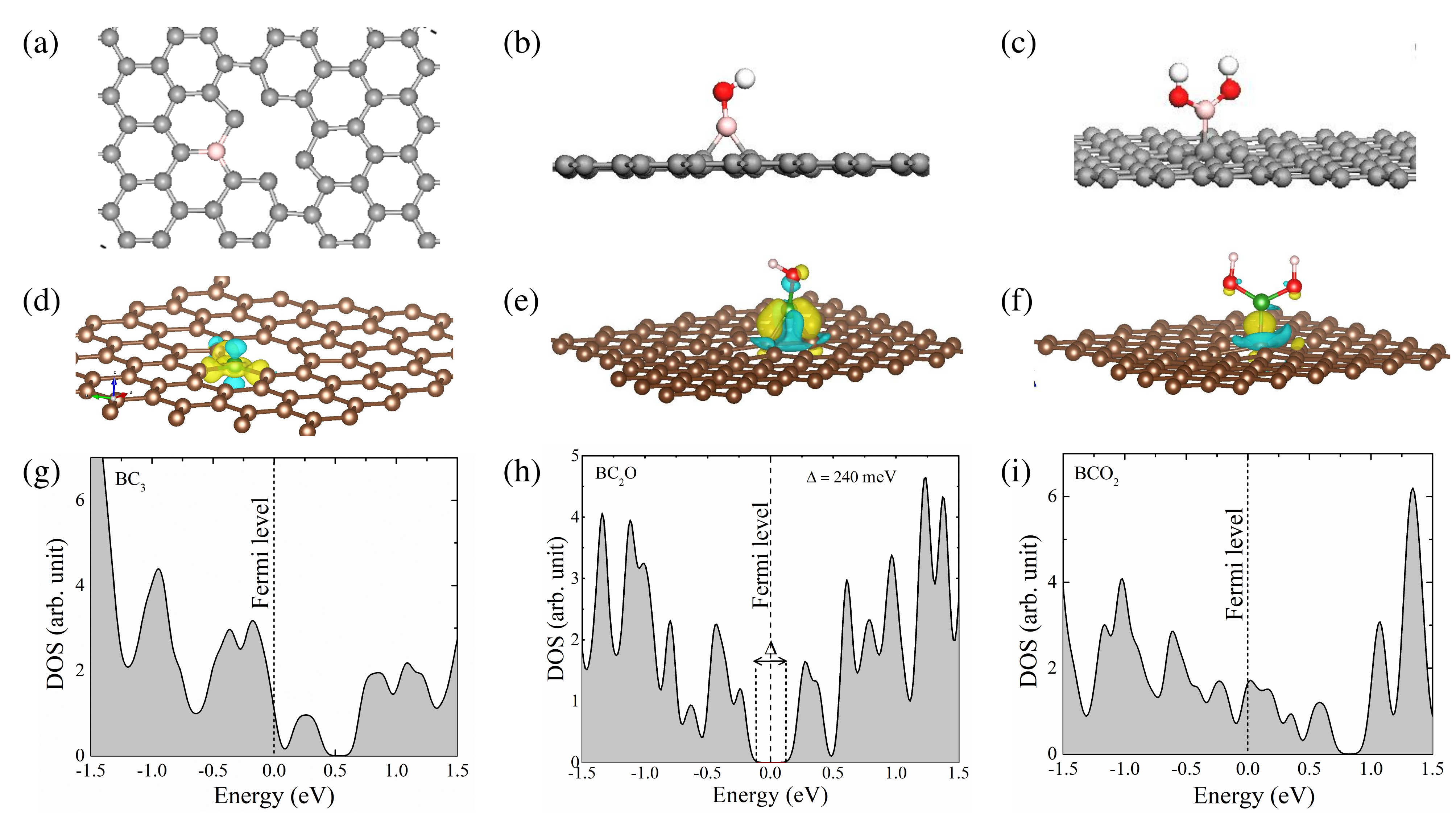}
\caption{Theoretical study: Optimized structure of boron doped supercell where boron atoms are configured as (a) BC$_3$, (b) BC$_2$O, and (c) BCO$_2$. Transfer of charges between in-plane carbon lattice and the (d) doped boron in plane, (e) BC$_2$O group, (f) BCO$_2$ group. Yellow shows the accumulation of electrons where cyan indicates deficiency of electrons after equilibrium. DOS vs. energy plot of (g) BC$_3$, (h) BC$_2$O, and (i) BCO$_2$ . $\Delta$ is the energy gap appeared at Fermi level in ground state.}
\end{figure*}

This work deals with a signature of the CO phase in non-metallic, boron-doped reduced graphene oxide where some boronic and borinic configuration of the boron atoms hold the interlayer positions along out-of-plane (z-axis) direction from lattice plane. Boron atoms are intentionally opted to incorporate between the graphene layers due to its light weight (small atomic mass, M) which is essential for the soft vibrations along the out-of-plane of lattice followed by an enhanced EPC. Generation of IL states associated with those out-of-plane doping configurations along with the transfer of charges between the IL states and graphene layer provide extra benefits towards obtaining such instability. Moreover, Dresselhaus predicted the superconducting state in boron doped graphene, commenting about unusual defect properties and localized electron density \cite{Kim2012}. These also act as regulating factors for enhancement EPC and hence for CO, as discussed later in this literature.

\section{\label{sec:level2}Materials and Methods}
\subsection{Materials}
\subsubsection{Synthesis of Graphene oxide}
All solvents and reagents used were of analytical grade and employed without further purification. Few-layer graphene oxide (GO) was synthesized from bulk graphite powder using the Modified Hummer’s method \cite{Cote2009}. In a typical synthesis, 1.0 g of graphite powder was added to a mixture of 45 ml concentrated sulfuric acid $(H_2SO_4, 98\%)$ and 5 ml phosphoric acid $(H_3PO_4)$ at room temperature. After stirring overnight, the mixture was cooled to $4^\circ C$ using an ice bath. Then, 6.0 g of potassium permanganate $(KMnO_4)$ was slowly added under continuous stirring while maintaining the temperature below $10^\circ C$. After 3 hours, 120 ml of distilled water was added, followed by 1 hour of continuous stirring. Subsequently, 8 ml of $30\%$ hydrogen peroxide $(H_2O_2)$ was introduced, and the solution was diluted by adding 250 ml of deionized water. The resulting yellowish precipitate was filtered and washed with $5\%$ HCl aqueous solution to remove residual metal ions. Further washing with deionized water was carried out until the pH reached approximately 6. The solution was then subjected to probe sonication at 500 W for 30 minutes to achieve better exfoliation.

\subsubsection{Synthesis of Boron-doped redueced graphene oxide}
To achieve our objective, we prepared several sample specimens of boron-doped reduced graphene oxide (BG) systems having different doping configurations of boron. To prepare BG samples, 50 mg of boron oxide $(B_2O_3)$ was added to 100 mL of an aqueous graphene oxide (GO) dispersion $(0.5 mg/mL)$ and ultrasonicated for 1 hour to achieve uniform mixing. The mixture was then dried at $50^\circ C$ over 24 hours. The dried product underwent thermal annealing at  $500^\circ C$, $700^\circ C$, $900^\circ C$, and $1000^\circ C$ for 1 hour each under a nitrogen atmosphere using a tube furnace. The resulting black powder was repeatedly washed with hot distilled water to eliminate any residual boron oxide and filtered multiple times. After drying, the final BG materials were collected. Samples were labeled BG500, BG700, BG900, and BG1000, corresponding to the respective annealing temperatures. \cite{Sarkar2021}.

\subsection{Experimental method}
  \subsubsection{XPS measurement}
	X$-$ray photoelectron spectra (XPS) for B1s peaks of BG samples were obtained by X$-$ray Photoelectron Spectrometer (SPECS, HSA 3500). Samples were kept under a monochromatic $Al-K_\alpha$ X-ray source of 1486.61 eV at ultra-high vacuum ($10^{-9}$ mbar) chamber. Highly resolved B1s spectra were deconvoluted with Lorentzian asymmetric line shape with tail damping, $LF(\alpha,\beta,w,m)$,  which is equivalent to asymptotic theoretical Doniach$-$Sunjic asymmetric line shape. Backgrounds were subtracted with Shirley algorithm. Contributions from each type of boron-carbon bonds were properly quantified using appropriate Relative Sensitivity factor (R.S.F.) value for boron as 0.482 to get accurate. High resolution scan for B1s were achieved by lowering the pass energy and regulating the step size, dwell time to count electrons that were coming to the hemispherical analyzer during measurement.
Binding energies and Full Width Half Maxima of all components found in B1s Peaks for all BG samples are provided in Supplementary table S1. Amounts of different doping configurations of boron atoms are given in Supplementary table S2. Deconvolution of B1s spectra for BG700 sample is shown in Supplementary figure S1.

\subsubsection{XRD measurement}
	Powder X-ray diffraction (XRD) patterns of the BG samples [Supplementary figure S2] were collected by Bruker D8 diffractometer with $Cu-K\alpha$ radiation at room temperature. An X-ray of wavelength,  $\lambda = 1.5404$ \AA $(Cu-K\alpha)$ was used for X-ray diffraction. Inter-layer distance along $(002)$ plane, $d_{002}$, for BG samples were calculated using the Bragg’s equation: $2d_{002} \sin{\theta} = n\lambda$, where $n=1$ for first order diffraction pattern and $\lambda$ is the wavelength of the applied X$-$ray.  Voigt functions were used to fit for the assessment of in$-$plane and crystallite size $(L_a)$ as shown in Supplementary figure S3. Employing deconvoluted parameters such as values of $2\theta$, Full width half maxima $(\beta)$ values were used to evaluate this crystallite size by applying Debye$-$Scherrer equation as, $ L =  K\lambda/(\beta \cos{\theta})$ , where, $K$ is Debye-Scherrer constant with the value of 1.84 for the in-plane crystallite of samples.

 \subsubsection{Raman Spectroscopy measurement}
 	Raman Spectra of all samples were characterized by Raman Spectrometer (WITEC, alpha 300 R) with an argon laser of wavelength, $\lambda = 532 nm$ and a 100X objective with 600 grooves/mm. All first and second order peaks for different vibrational modes present in the samples were deconvoluted using nine Lorentzian functions as presented in Supplementary figure S4. Total integrated area of first order D and G peaks were considered to calculate the  $I_D/I_G$  ratio. To get  $I_2D/I_2D'$   ratio for equation $5$, integrated area of 2D and 2D’ peaks were measured. 
Crystallite size $(L_a)$ and defect length $(L_D)$ of the samples, mentioned in Figure 2(e), were calculated using the equations \cite{Cançado2006,Cançado2011},
\begin{equation} 
L_a (nm) =\frac{(2.4 \times 10^{-10}) \lambda^4}{I_D/I_G}
\end{equation}
\begin{equation} 
{L_d}^2 (nm^2) =\frac{[(1.8 \pm 0.5) \times 10^{-9}] \lambda^4}{I_D/I_G}
\end{equation}
In figure 2(d) and 2 (f), measurements of the values strain and doping are performed through Raman analyses using ref. [\cite{Sarkar2021}; figure 9]. Individual contributions of strain and doping in shifting of the G and 2D bands for each sample were estimated by deriving the component of the “shift” along tensile strain and hole doping axes respectively. Lee et al. conjectured about the contributions of the strain and doping in band shifts, as mapped with two axes, could be realized through vector method \cite{Lee2012}. 

\subsubsection{Electrical Properties measurement}
$0.8 mm$ thick circular pellets of the BG samples having $2.8 mm$ of radius were prepared to measure the temperature dependent resistance by applying current $(I)$ through the standard four terminal $(1\ mm\ gap)$ electrode system in Cryogenics $(UK)$ Physical Property Measurement System from $200K$ to $4.2K$. Terminals are constructed by silver $($Ag$)$ deposition on pellet sample and the gap between terminals are measured as around $1.0$ mm by a traveling microscope having least count of $0.01mm$. A schematic diagram of terminal connections is provided in supplementary information $($figure S5$)$. In this study, static resistance $(R=V/I)$ is used to evaluate the temperature--dependent electrical behavior of semiconducting graphene. Despite the nonlinear $I-V$ characteristics often observed in doped or oxidized graphene, static resistance effectively captures the material’s overall transport behavior, which is governed by temperature-dependent parameters such as carrier concentration and mobility. In contrast, dynamic resistance $(r=dV/dI)$ varies with the operating point and is unsuitable for tracking bulk thermal effects. Moreover, models like variable range hopping $($VRH$)$ rely on static resistance, making it the standard and physically meaningful choice for such analyses. 

\subsubsection{Low temperature Raman Spectroscopy measurement}
Raman measurements were performed using a 532 nm continuous-wave laser as the excitation source. The laser power was 1.7 mW measured prior to the objective lens. The excitation beam was focused onto the sample using a 50× objective lens with a numerical aperture (NA) of 0.5.
Samples were mounted inside a closed-cycle helium cryostat equipped with a top optical window, allowing temperature-dependent measurements under high vacuum conditions $(<10^{-4}$ $Torr)$. Raman spectra were acquired in steady-state mode using a Renishaw spectrometer equipped with a 1800 lines/mm grating and a silicon CCD detector. All measurements were conducted under vacuum to minimize thermal and environmental noise and to ensure thermal equilibrium at low temperatures.

\subsection{Theoretical Methods}
In the current work, the first principles calculations were performed within the scope of density functional theory (DFT) using the Vienna ab$-$initio simulation package (VASP) \cite{Kresse1993,Kresse1994,Kresse1996}. All the calculations were carried out with the projector augmented wave (PAW) \cite{Blochl1994} method. The exchange$-$correlation terms were treated at the level of generalized gradient approximation (GGA) by means of the Perdew$-$Burke$-$Ernzerhof (PBE) \cite{Perdew} functional. The structural optimizations were allowed to continue until the difference in the energy of the system in two consecutive iteration steps reached below $10^{-5} eV/ atom$. The calculations were performed at the $\Gamma$ point with an energy cut-off value of 500 eV. The Brillouin zone integration was done for a k$-$point mesh of $(4\times4\times1)$. A vacuum slab of length 20 \AA was implemented in the direction perpendicular to the two$-$dimensional surfaces to ward$-$off spurious interaction between the periodic images. The DFT+D2 (Grimme’s) method \cite{Grimme2006} was utilized to take into account the effect of the dispersive forces. In order to evaluate the band gap value with higher accuracy, the Heyd$-$Scuseria$-$Ernzerhof (HSE06) hybrid functional \cite{Krukau2006} was used during the density of states (DOS) calculations [Figure 1], whereas the band structures and their orbital decomposed versions were computed within GGA$-$PBE level calculation in order to save computational time [Supplementary Figure S7]. 

\section{\label{sec:level3} Results and Discussion}
\subsection{DFT analyses}
We optimized our BG samples with different doping configurations for ab-initio study. Fig 1(a), 1(b) and 1(c) show the optimized structure of doped graphene consisting of the in-plane BC$_{3}$, out-of-plane BC$_2$O and BCO$_2$ groups respectively. Choice of the di-vacancy defect is made to emulate the real system as prepared by annealing treatment at high temperatures \cite{Banhart2010,Tsetseris2014}. Fig. 1(e) and 1(f) exhibit localization of the charges at out-of-plane dopant sites as electrons are transferred to the out-of-plane boron groups from the lattice plane. Enhancement of the population of the electrons at the IL region is the most suitable arrangement for the excitation of those electrons to the anti-bonding state easily. Some metal intercalated graphite compounds procure the SC states using this charge transfer process \cite{Csnyi2005}. On the other hand, electrons are more localized around boron itself for the in-plane doping as shown in fig 1(d). There is another viewpoint for the charge transfer process as the lacking of charge from the lattice for the out-of-plane doping leaves the system more hole doped, we provide a different approach explained later in this work.

\begin{figure*}
\includegraphics[width=\textwidth]{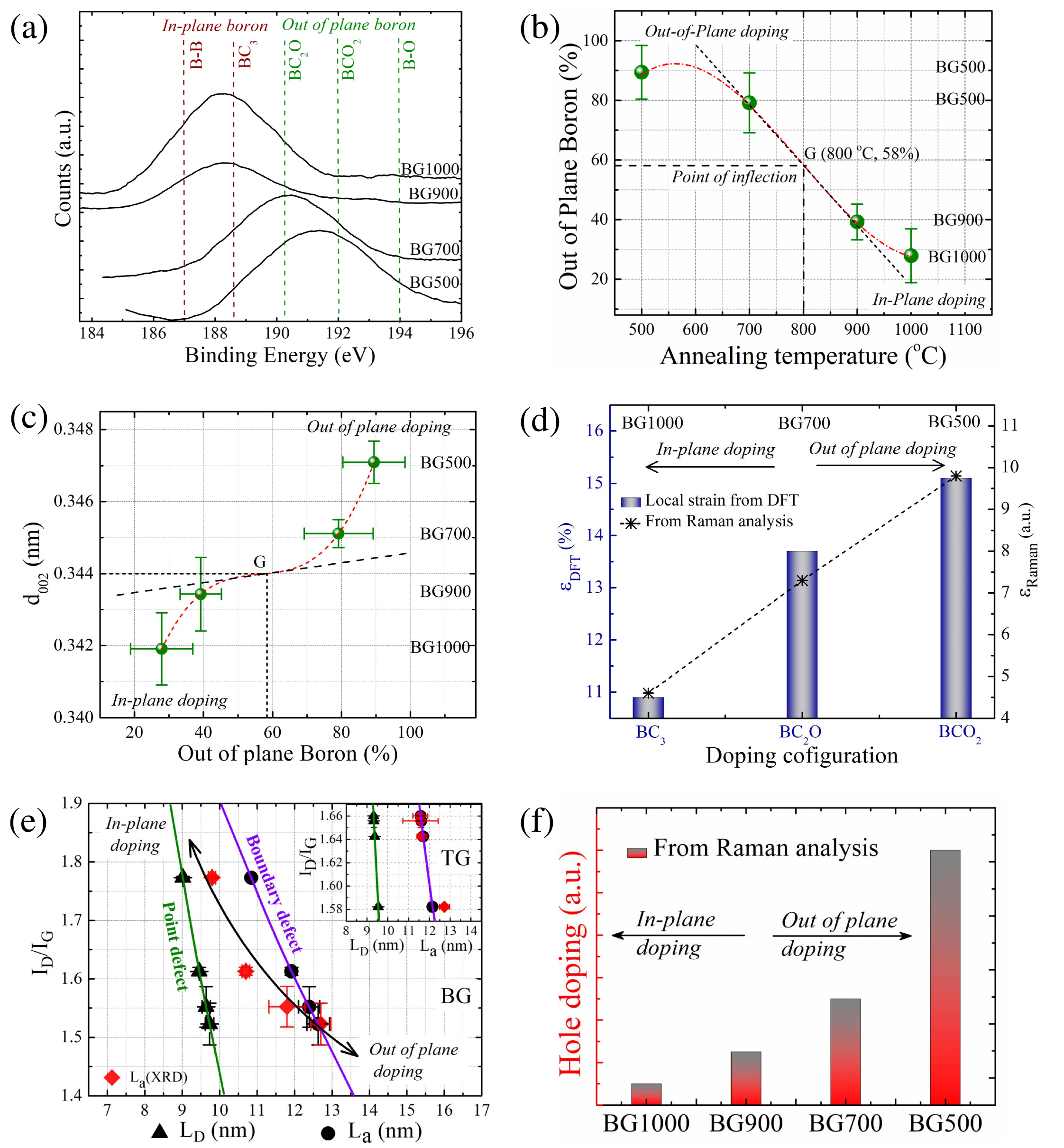}
\caption{Manipulating factors of BG samples in order to achieve CO state: (a) XPS study of all BG samples. (b) Real measures of out of plane boron and in-plane boron in BG samples. (c) Variation of interlayer distances with the amounts of out-of-plane groups present in BG samples. (d) Strain produced due to inclusion of differently configured boron groups in graphene lattice. Bars represent the result obtained from DFT, and stars demonstrate the strain amounts as realized from Fig. 9 in ref. 35. (e) Types of defects produced in graphene lattice as a result of boron doping. $I_D/I_G$ ratios as a function of experimentally obtained in-plane crystallite sizes, $L_a$ (XRD), of BG samples considering T-K relation and modified-T-K relation for boundary (characterized by crystallite size, $L_a$) and point defects (characterized by defect length, $L_D$) respectively [Supporting info.]. (f) Degree of hole hoping in all BG samples measuring from the Raman analysis [\cite{Sarkar2021}; Fig. 9].}
\end{figure*}

Now, to find the gap in the valance band owing to the energy reduction of the periodic density of charges, the total density of states (DOS) of all the doped systems are shown in Fig. 1(g), 1(h) and 1(i) respectively. For the in-plane doping of the boron, the Fermi level shifts down inside the valance states by an amount of 470 meV, as seen in fig 1(g). This shift clearly indicates hole doping due to boron inclusion in the graphene lattice. The gap between the conduction band (CB) and the valance band (VB) is found here to be about 90 meV. This gap is expected here due to the di-vacancy of graphene as well as the defect present in the lattice and the interstitial sites of graphene. On the extreme side, for BCO$_2$ configurations, this shift of the Fermi level to the VB is much larger and this shift of the Fermi level from the top of the valance band, simply designated as the valance band maxima (VBM), is about 770 meV. Hence much higher percentage of the hole doping is expected here, which also substantiate the experimental results. Bandgap for this structure is also increased to 110 meV as compared to the BC$_3$ configuration. Now, for the BC$_2$O configurations, instead of the shifting of the Fermi level towards the valance band, a wide bandgap of about 240 meV opened up at the Fermi level. Hence, the band separation at the Fermi level followed by a divergence of density of states near the saddle points emphasize the possibility of charge ordering below transition temperature. In addition, we observe Van Hove Singularity (VHS) slightly away from the Fermi level ~0.5eV for this BC$_2$O configuration as seen in figure 1(h). Furthermore, a flat like band near the Fermi level is also observed for the BCO$_2$ configuration as well [Fig. S7; Supporting Info.]. Other interfering factors which could enhance EPC more efficiently in graphene is the density of states at Fermi level, introduced as $N(E_F$) in equation (1). After realizing the shift of $E_F$ into the VB, from figure 1, edge of the VB is further realized by inspecting the electronic band structure for the corresponding out-of-plane boron configurations [supplementary information]. 

\subsection{Doping information}
Fig. 2(a) depicts the XPS results obtained for all of the BG samples. Peaks associating with the other out-of-plane configurations of boron atoms are found above B.E. 190 eV. As the synthesis temperature goes to  $1000^\circ C $, most of these out-of-plane groups are removed and thereby substitutional doping (mostly BC$_3$) in-plane of the lattice is ensued in majority as depicted in Fig. 2(a).We quantified the in-plane and the out-of-plane boron groups present in the graphene samples physically by deconvoluting the B1s peak for each sample [Fig. S1; Supporting Info.]. Evolution of the doping configuration from the out-of-plane to the in-plane is shown in Fig. 2(b). Out-of-plane boron groups own maximum occupancy about 90\% of the total doping for then BG500, which then falls gradually to 30\% for the BG1000 sample [Table S2; Supporting Info.]. Amounts in percentages are fit with the cubic function that show the two distinct stages, by means of doping, separated by an inflection point (Fig. 2b). At this point, the out-of-plane boron groups holds around 60\% occupancy, which is found to be the mid-value of the entire range of the doping variation. Hence this current state of affairs distinguishes two different stages modulated by the in-plane (I/P) and the out-of-plane (O/P) boron doping in graphene systems. Moreover, the degree of hole hoping in all BG samples measured by using the Raman analysis from our previous work[\cite{Sarkar2021}; Fig. 9]. Hole doping gradually increases with increasing O/P boron groups in graphene and reaches to maximum in BG500 as depicted in figure 2(f) Here, some contrastive physical, electrical, vibrational properties at these two stages are found to be either collectively manipulating the electronic instabilities or be a by-product of this phase transition. 
\begin{figure*}
\includegraphics[width=\textwidth]{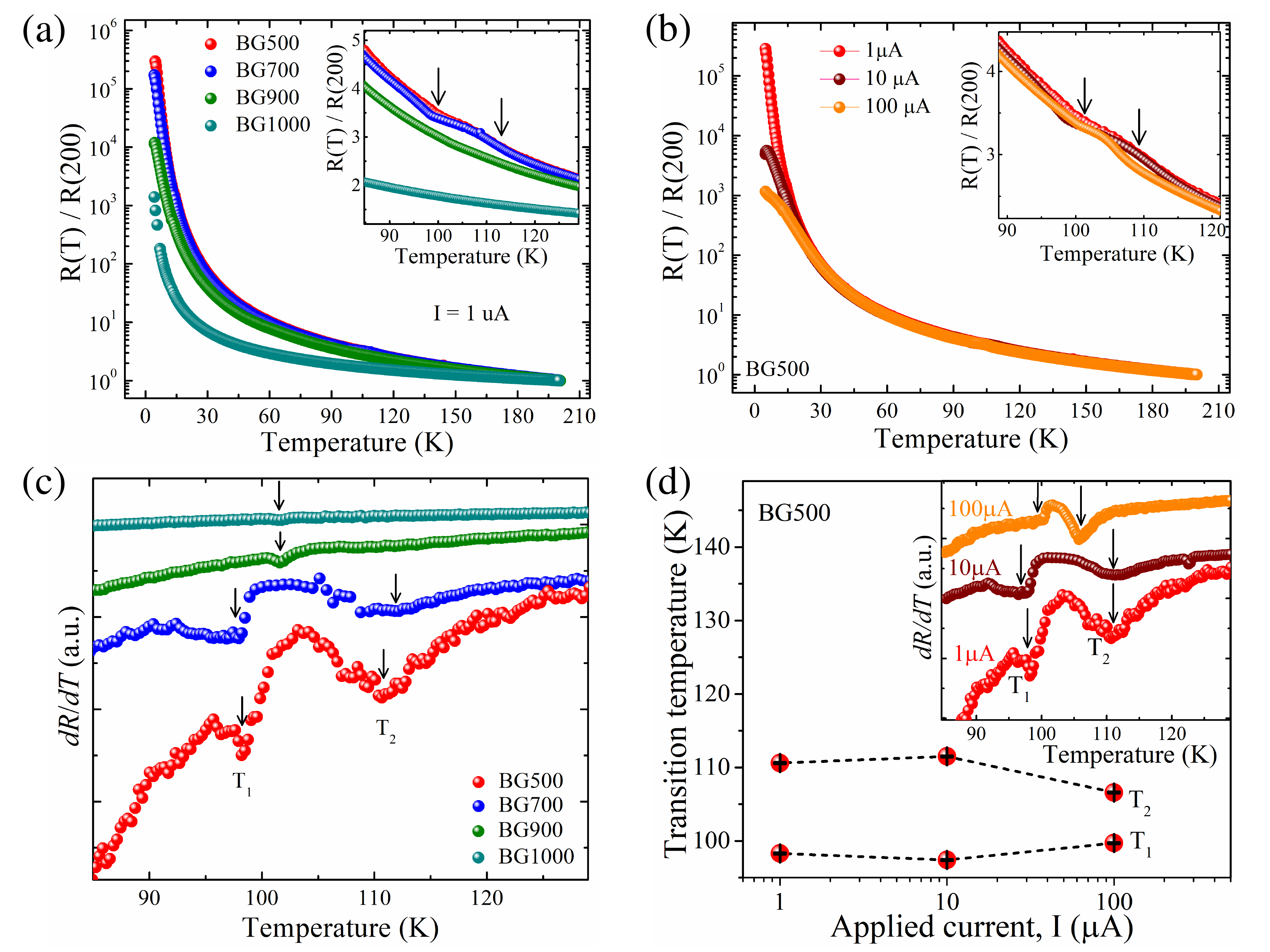}
\caption{Temperature dependent resistances (R) of all BG samples under a bias current $I = 1\mu A$. (b) Temperature dependent resistances of BG500 sample under various bias currents (c) Corresponding first derivative of resistance curve for all BG samples at bias current $I = 1\mu A$. (d) Influence of applied bias current on transition temperatures, T$_{1}$ and T$_{2}$. Corresponding first derivative of resistance curve for BG500 sample at different bias currents (inset).}
\end{figure*}

\subsection{Structural analyses}
First of all, the IL distances of these doped graphene samples, specified by $d_{002}$ in Fig. 2(c), demonstrates the increasing trend of the IL gap with increase in doping occupancy for the O/P sites. $d_{002}$ for the BG samples that are obtained from the X-ray diffraction patterns [Fig. S2; Supporting Info.]. BG1000 has the shortest IL distance among all the BGs due to several factors like removal of the O/P boron groups, deficiency of the $\pi$-electron densities, van der Waals forces etc. which were clearly discussed in ref. \cite{Sarkar2021}. On the other hand, BG500 has the maximum IL distance where accumulation of the electrons in IL region definitely modifies the vibrational states in achieving CO states. Another stimulating factor for the acquisition of the CO phase in two-dimensional materials is tensile strain produced along the I/P direction of the lattice \cite{Gao2018,Wei2017}. Some two-dimensional TMDs such as 2H-NbSe$_2$ \cite{Gao2018} 1T-VSe$_2$ \cite{Zhang2017} and rare-earth ditelluride like CeTe$_2$ \cite{Sharma2020} have shown their CO transition where tensile strain directly helps in the charge ordering. Tensile strain, particularly uniaxial one (as biaxial strains have little contribution to it) enlarges the momentum separation $q_{CO}$ in Fermi pockets around the $\Gamma$ point by softening the phonon energies and thereby altering the carrier hopping parameter \cite{Gao2018}. It is now established that the boron doping in hexagonal carbon lattice induces tensile strain \cite{Kim2012,Lee2012}. Moreover, negative thermal coefficient of graphene is also a considerable factor at low temperatures, as it stretches the carbon rings during the cooling down of the temperature \cite{Yoon2011}. Our earlier literature gives an estimation of the tensile strain due to the boron groups in BG samples [Ref. \cite{Sarkar2021}; Fig. 9]. Fig. 2(d) depicts that O/P boron groups develop more strain than I/P boron groups inside the graphene lattice which evidently assists the CO phase at lower temperatures. To endorse the  experimental strain analysis, ab-initio simulations are investigated, as described in Fig. 2(d), to find effective strain induced in all BG samples. Theoretically, amounts of strain produced in BG samples due to boron inclusion were evaluated by measuring the B$-$C bond length. Length of different bonds for three standard boron configurations are given in Supplementary table S3. Theoretical strain for a particular boron groups are compared with the strain in experimental BG systems associated with the most populated boron groups in the system. Since the O/P boron groups take the utmost responsibility to produce tensile strain in BG700 and BG500 lattice as compared to the other BG systems. 

\begin{figure*}
\includegraphics[width=\textwidth]{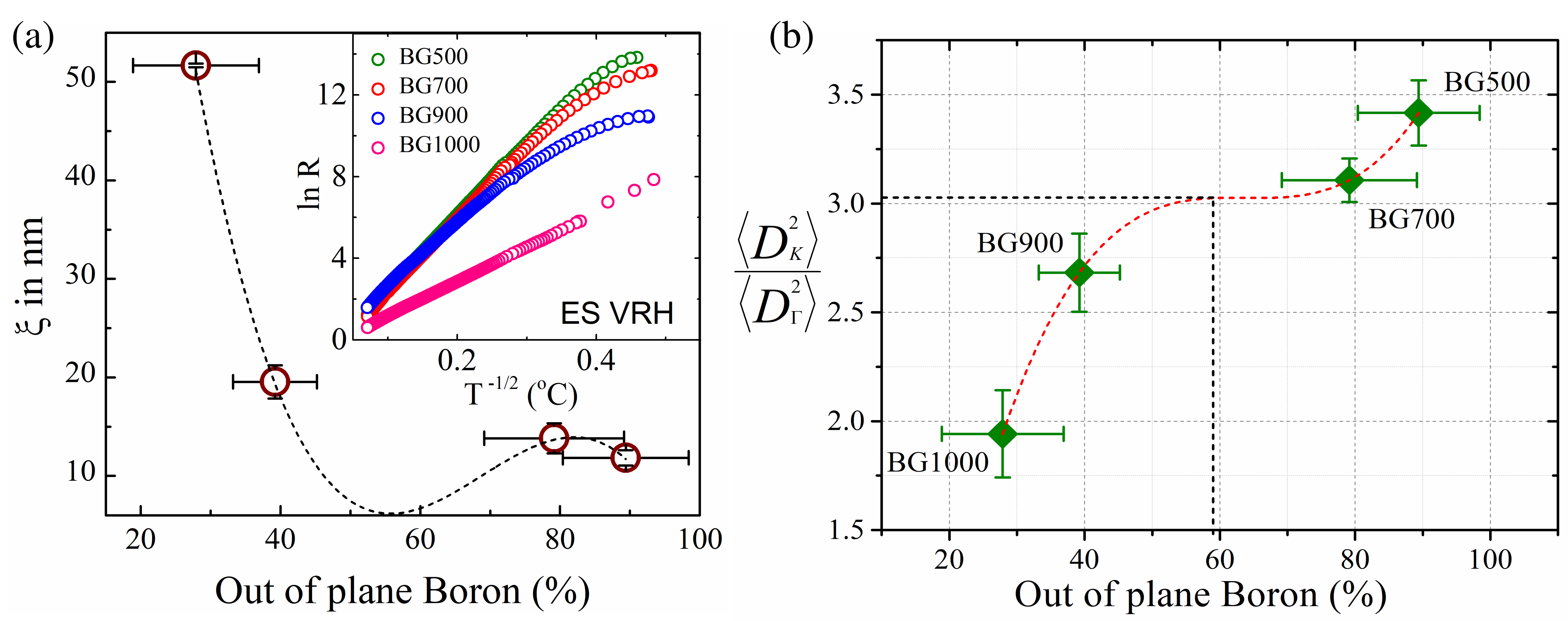}
\caption{(a) Variation of localization length ($\xi$) with different amounts of O$/$P boron, (b) Ratio of the average of the square of EPC parameters, averaged on the Fermi surface, associated with the K and $\Gamma$ points, represented as $\langle D_K^2 \rangle/\langle D_\Gamma^2\rangle$  vs. out-of-plane boron percentage.}
\end{figure*}
\subsection{Role of defects}
Furthermore, the role of defects in CO phase transition are unexplained till date. Weitering et al. clearly describe how point defects acts as nucleation centers for the CO phase in germanium lattice \cite{Weitering1999}.  Fig. 2(e) reveals the nature of defects occurring in our boron doped graphene samples. After calculating the $I_D$/$I_G$  values from the deconvoluted Raman spectra of BG samples [Fig. S4; Supporting Info.], two distinct reference lines for point and boundary defects are drawn using the modified T-K relation \cite{Canado2011}. Point defects and boundary defects are characterized by their defect length ($L_D$) and crystallite size ($L_a$) respectively in the figure. For two dimensional lattices, both of these parameters reveal similar range of crystallinity. Therefore, an experimental measures of crystallite sizes are cumulated by analyzing the XRD peaks along in-plane (001) direction [Fig. S3; Supporting Info.]. It becomes clear that the variation of the crystallite sizes in the BG samples indicate that the O/P doping creates the boundary type defects inside the lattice whereas, I/P substitution of boron creates the point type defects in the same. On the other hand, thermally treated undoped graphene (TG) samples, are prone to have boundary defects for its different functional groups attached in the out-of-plane configuration [Fig. 2(e); inset]. However, we believe that the boundary defects in our O/P doped system divides the whole lattice into several grains and the possibility of incommensurate CO (IC-CO) arises since commensurable phase transition requires lesser dimension of defects \cite{Kogar2017a,Yan2017}. Absence of loop for the heating and cooling cycle at low temperature transport measurements also indicate a transition is likely to be from normal to IC-CO states in BG500 and BG700 samples \cite{Baraghani2022,Sipos2008,Stojchevska2014}. Thus, in present scenario, all those factors like, tensile strain, hole doping, line defects etc. are directly or indirectly trigger the O/P boron doped BG system to have greater EPC that allow for the charge ordering and periodic modulation of electron density in one dimension.

\subsection{Electrical transport mechanism}
Now to observe the charge transport phenomenon particularly in the low temperature regime, resistance vs. temperature ($R-T$) measurements were performed for all the BG samples [supporting Info.]. Fig. 3(a) reveals the semiconducting nature of all BG samples with long range disorder. Resistance values are normalized at 200K to compare the order of magnitudes and their derivatives ($dR$/$dT$) for each sample as well as for each bias current ($I$).  An observable kink near 100 K can be identified for both the BG500 and BG700 samples under applied $I = 1\mu A$ (inset: fig. 3(a)). These small elevations in resistance values just above 100K are expected due to reconstruction of in-plane superlattice leading to a CO phase transition in the BG system. Transition from normal to CO phase are more prominent in the derivative plots for the BG samples in fig. 3(c). Interestingly, in $dR$/$dT$ representation, two points of inflections clearly infers that the formation of superlattice takes place by two sets of transitions in samples. Two distinct temperatures separated by nearly 10K are designated as T$_{1}$ and T$_{2}$ in Fig. 3(c). It is obviously unconventional to have such pair of minima in the way of achieving CO phase transition in other TMDs, but still for few-layer, boron-doped graphene comprised with large amounts of defects it can be predicted due to few dissimilar aspects compared to normal TMDs. 
\begin{figure*}
\includegraphics[width=\textwidth]{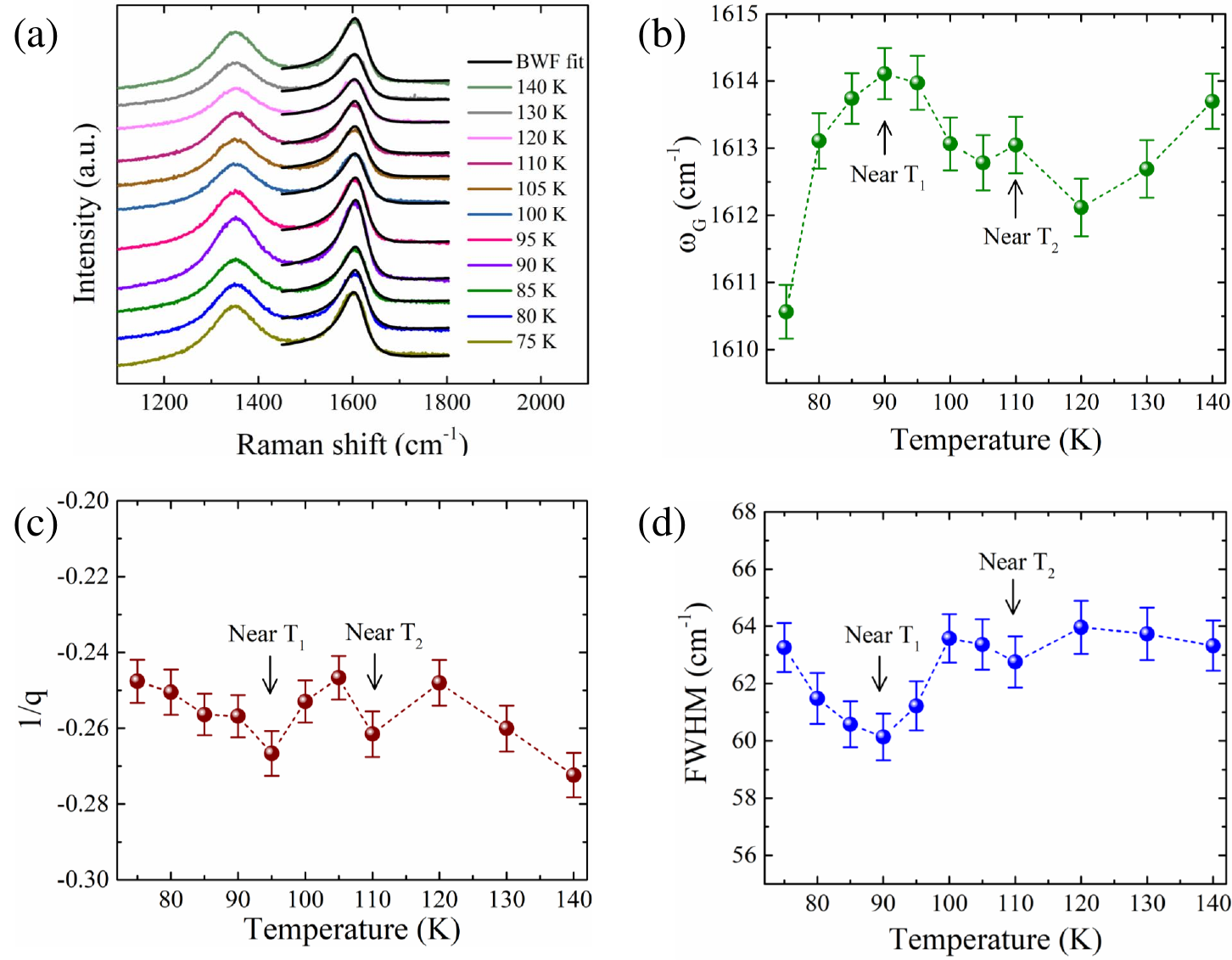}
\caption{(a) Temperature dependent Raman spectra for BG500. G peaks are fitted with Breit-wigner-Fano lineshapes. Solid black lines are fitted curve (b) Variation of G peak position at different temperatures. (c) 1/q, inverse of assymetry factor versus temperature plot. (d) Width of G peak versus temperature plot.}
\end{figure*}
Effect of the applied electric field in achieved CO phase set another perspective to the understanding of this collective phenomena. Different $R-T$ curves and first order derivative with respect to the temperature in absolute scale are obtained for BG500 at $I = 1\mu A$, $10\mu A$ and $100\mu A$.These measurements were conducted to probe the influence of the electric field in CO phase transition, as depicted in Fig. 3(b) and 3(d) respectively. As the intensity of applied electric field increases, destruction of the CO phase occurs. It is possible that pinning of the electric field strengthens the coulombic interactions substantially, and thereby harden the vibration of the phonons by significant amount. This presumably yields in vanishing of the CO phase because of the weaker EPC. Similar observation is reported by Balandin group  \cite{Geremew2019}, where higher electric field redirects the sample 1T-TaS$_2$ from IC-CO to metallic state. For BG500 sample, variation of the onset points for CO transition by altering the applied bias field are shown in Fig. 3(d), which delineates the evolutions of both T$_{1}$ and T$_{2}$.We observe that two transition points have a clear tendency to a single point of convergence suggesting the destruction of CO phase with increasing electric field. Some rare earth tritellurides holding two CO transition points previously depicted this type of shrinkage of CO signatures under increasing pressure \cite{Zocco2015,Ru2008}. 
 To identify a suitable mechanism for temperature-dependent transport in our samples, we analyze the resistance vs. temperature ($R$–$T$) data using three established models: the Arrhenius model~\cite{Ci2010}, the variable range hopping (VRH) model by Mott~\cite{Mott1969, Mott1978, Mott1979}, and the Efros–Shklovskii (ES–VRH) model~\cite{Efros1975}. Among these, the ES–VRH model best describes the carrier transport in boron-doped graphene (BG) samples. Accordingly, the $R$–$T$ data for all samples are fitted [Figure~4(a), inset] using the following equation,
\begin{equation} 
R(T) = R(0) exp\left[\left(\frac{T_{ES}}{T}\right)^{\frac{1}{2}}\right]
\end{equation}
, where $R(T)$ is the resistance of the samples at $T K$, $R(0)$ is a pre-exponential factor and $T_{ES}$ is the characteristics temperature. Localization length $\left(\xi\right)$, as described in figure 4(a), is another inferential phenomenon linked with VRH mechanism of charge conduction, and this can be derived from $T_{ES}$  as \cite{Joung2012},
\begin{equation} 
\xi = \frac{\beta_1e^2}{k_{\beta}T_{ES}\epsilon'}
\end{equation}
, where $\beta_1 = 6.2$ for $3D$ ES-VRH and $\beta_1 = 2.8$  for $2D$ ES-VRH, $k_\beta$ is Boltzmann constant and $\epsilon'$ is the permittivity of of the boron doped graphene system. 
Fig. 4(a) indicates that the BG samples with more out-of-plane boron groups are having shorter localization length,$\xi$. Localization of electron wave function in BG500 and BG700 can be understood as a by-product of the enhanced EPC. Hence, the EPC terms definitely plays the utmost role in the genesis of CO phase in BG500 and BG700. 

\subsection{Origin of charge ordered state}
To understand the origin of this CO phase in BG, two most discussed mechanisms are brought into this picture. Firstly, Fermi surface nesting (FSN) which is characterized by a wave vector $q_{CO}$ that connects the Fermi contours. Secondly, momentum dependence of electron-phonon coupling (EPC) matrix induces CO phases as observed in case of 2H-NbSe$_2$. There is another mechanism of CO, mostly found in some cuprates, termed as unconventional density wave \cite{Fujita2014}. Unlike NbSe$_2$, TaS$_2$, TaSe$_2$ etc. CO in cuprates cannot be explained by FSN or $q$-depended EPC, but rather influences of anti-ferromagnetism and coulomb interaction are anticipated \cite{Zhu2015}. As present work deals with mainly light atom doped graphene system, the possibility of unconventional charge density waves can be discarded. Possibility of FSN is mostly seen in one dimensional lattice \cite{Johannes2008} or some TMDs like VSe$_2$ etc. \cite{Duvjir2018}. q-depended EPC involves an inelastic scattering interaction of electrons in the lattice. Thus, to assess the momentum dependency and possibility of the EPC at higher symmetry points, second order Raman peaks 2D and 2D' originating from the interaction of the phonons at $K$ and $\Gamma$ points are precisely examined [Supporting Info.] \cite{Attaccalite2010,Alzina2010}. Fig. 4(b) dictates that the EPC near $K$-point is amplified in graphene containing the O/P boron groups. 
	
    To gauge the strength of electron--phonon coupling $\left(EPC\right)$, we put more focus on the deformation potential, which is strictly related to the EPC strength, as given in equation (1).  Now, the square of the average of the deformation potentials, precisely related to the vibration of phonons near higher symmetry points, are directly proportional to the integrated peak intensity of $2D$ and $2D'$ peaks as they are generated after scattering with phonons near $K$ and $\Gamma$ points. Rise of $2D$ and $2D'$ peaks consists of fully-resonant four-step process, where photoexcited electron-hole $(e-h)$ pairs undergo intra-valley and inter-valley scattering with phonons. Now, the rate of overall inelastic scattering $(\gamma)$ of $e-h$ pairs includes the electron-electron/hole-hole $(e-e)$ or $(h-h)$ scattering rate $\left(\gamma_{e-e}\right)$ or $\left(\gamma_{h-h}\right)$ and the electron-phonon $\left(e-ph\right)$ scattering rate $\left(\gamma_{e-ph}\right)$, among which $\gamma_{e-e}$ or  $\gamma_{h-h}$ depends linearly with Fermi energy level. The $e-ph$ scattering contribution comes from the phonons at $K$ and $\Gamma$ points, assigned as $\gamma_K$ and $\gamma_\Gamma$   respectively. Hence the integrated intensities of $2D$ and $2D'$ can be expressed in terms of scattering rate as: \[I(2D) = 2C \left( \frac{\gamma_K}{\gamma} \right)^2, \quad I(2D') = C \left( \frac{\gamma_\Gamma}{\gamma} \right)^2,\], where $C$ is a proportionality constant. Again, the scattering rate due to the $K$ and $\Gamma$ phonons are directly proportional to the corresponding dimensionless EPC parameters termed as $\lambda_K$ and $\lambda_\Gamma$. Hence, the ratio of EPC parameters associated with phonons at $K$ and $\Gamma$ can be obtained from the ratio of integrated Raman intensities:
\[
\frac{\lambda_K}{\lambda_\Gamma} \propto \frac{I(2D)}{I(2D')}.
\] \cite{Alzina2010}.  Taking all these relations into account, $\lambda_K/\lambda_\Gamma$ can be formulated as,
    
\begin{equation} 
\frac{\lambda_K}{\lambda_\Gamma} = \frac{1}{\sqrt{2}} \sqrt{\frac{I\left(2D\right)}{I\left(2D'\right)}} \frac{E_L-2\omega_\Gamma}{E_L-2\omega_K}
\end{equation}
, where $E_L$ is the energy of incident photon, $\omega_K$   and $\omega_\Gamma$  are the frequency of scattered photon at $K$ and $\Gamma$ point respectively. Ratio of the average of the square of deformation potential, averaged on the Fermi surface, associated with the $K$ and $\Gamma$ points can also be calculated as,
\begin{equation} 
\frac{{\langle D_K^2 \rangle}_F}{{\langle D_\Gamma^2 \rangle}_F} = \sqrt{2} \sqrt{\frac{I\left(2D\right)}{I\left(2D'\right)}} \frac{\omega_K\left(E_L-2\omega_\Gamma\right)}{\omega_\Gamma \left(E_L-2\omega_K\right)}
\end{equation}
Where, ${\langle D_K^2 \rangle}_F$ and ${\langle D_\Gamma^2 \rangle}_F$ are the deformation potentials allied with $K$ and $\Gamma$ phonons \cite{Attaccalite2010}. However, some authors recognized these as the EPC matrix elements for the respective higher symmetry points \cite{Alzina2010}. Here, considering the very straightforward dependency of deformation potential to EPC strength as shown in eq. (1) in the main article, we simply call it as EPC parameters.   Above calculations reveals that the EPC near $K$ point is higher compared to $\Gamma$ point for all the BG samples, described in figure 4(b). Moreover, Considering aforementioned the coulombic interactions near to the Fermi level and the isotropic nature of the system, these samples are best fit [Fig. S6; Supporting Info.] to the Efros-Shklovskii variable range hopping (ES-VRH) model \cite{Efros1975,Joung2012} of conduction of charge carriers.

\begin{figure*}
\includegraphics[width=9cm]{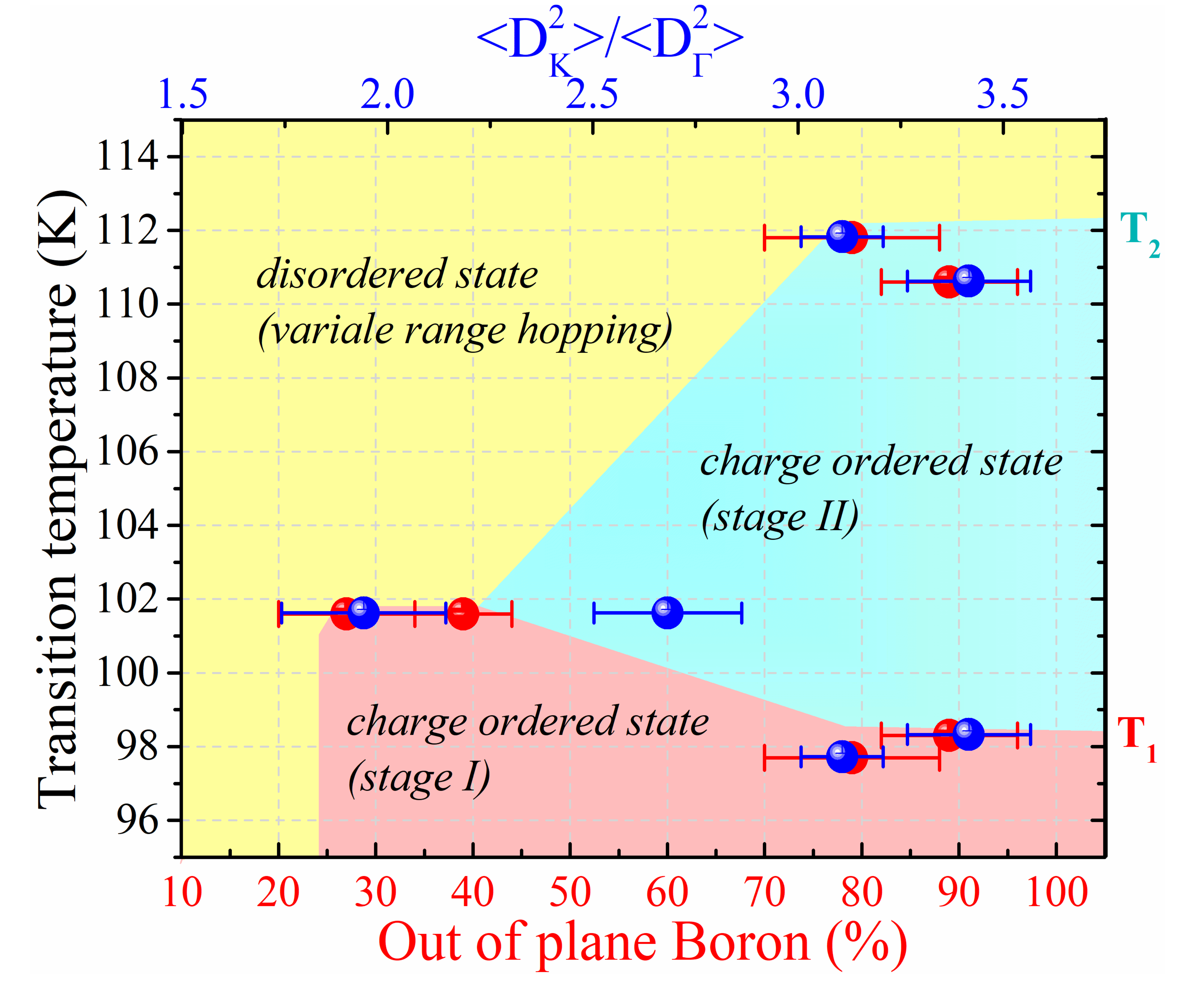}
\caption{Transition temperatures vs. percentage of O/P boron phase diagram: Elucidating the effect of O/P doping and associated EPC parameter  ($\langle D_K^2 \rangle/\langle D_\Gamma^2\rangle$ )  [as calculated in Fig. 4(b)] on transition temperatures, T$_{1}$ and T$_{2}$ values for four out-of-plane boron doping percentage.}
\end{figure*}
Starting from a non-metallic ground state, the first CO transition occurs near $T_{1} = 100K$ are obtained by modulation of charge periodicity as traditionally occurs. Now, a possible reason for subsequent transition, at $T_{2} = 110K$, is anticipated as the sufficient incorporation of O/P boron can modify the surface of graphene layer by generating a pseudo-periodicity of superficial atoms as boron has a greater self-diffusive property in carbon lattice at higher temperature \cite{SuarezMartinez2007}. It is noteworthy, that the BG900 and BG 1000 samples containing mostly in-plane boron atoms have single transition temperature as there is no such surface boron which can contribute to this pseudo-periodicity. Fluctuations of onset transition temperatures for differently doped BG systems are elucidated in Fig. 5. This phase depicts the final result precisely, as the signatures of CO phases are found to be more profound in the samples with greater number of O/P groups. Strength of EPC induced by out-of-plane vibrations is also affecting the transition from insulating to CO and CO to semiconducting state in similar fashion as represented this figure. Therefore, the significance of O/P boron groups as a modulating factor in order to attain strongly correlated system of particles followed by the lattice reconstruction at this particular temperature ($T_1$) is understood as predicted from theoretical analyses. 

\subsection{Temperature depended Raman Spectra analyses}
To further support our findings from low temperature dependent transport measurements, we employ complementary low temperature Raman spectroscopy. In figure 5 (a), the stacked plot representation of low temperature Raman data for the BG500 sample is shown from 75 – 140 K, i.e., the main temperature region concerning charge ordering and focusing on the D, G bands.  Previously it was reported that splitting of the G band occurs due to lattice reconstruction in multilayer graphene under applied electric field \cite{Long2016}. In this case, we do not observe any drastic changes to the Raman spectroscopy. However, on closer inspection we find minute features reminiscent of charge ordering after fitting the G band with Breit-Wigner-Fano line shape. The details of the fitting and its parameters are given in the supplementary information.
The use of the BWF line shape fitting is appropriate in this regard due to the highly asymmetric nature of the G band in our sample. In figure 5 (b, c, d), we showcase the peak position ($\omega_G$) , coupling strength $(1/q)$  and full width half maxima (FWHM) dispersion within the temperature range 75 – 140 K. It is interesting to observe complementary peak and dip like features in $\omega_G$ , FWHM dispersions to the designated $T_1$,$T_2$ charge ordering points from previous transport measurements (figure 3). This complementary effect is strongly indicative of charge ordering in our BG500 sample. Furthermore, the most interesting aspect is found in the $(1/q)$ where we clearly observe dips in the dispersion corresponding to the charge ordering temperatures $T_1$,$T_2$. It is well-known that existence of Fano line shape points to robust evidence of EPC which is the interference between electronic continuum states and discrete phonon mode. Moreover, the sign of the coupling strength $(1/q)$ designate the location of the charge continuum, i.e, conduction or valence band. Specifically, negative $(1/q)$ values are typically associated with hole doping (p-type behavior), while positive values indicate electron doping (n-type). This distinction arises from the phase relationship of the interfering pathways in Fano resonance: hole doping shifts the Fermi level below the Dirac point, modifying the electron–phonon coupling in such a way that the resulting G band asymmetry is skewed toward lower energies, yielding a negative 
$(1/q)$ This behavior has been experimentally demonstrated in electrostatically gated graphene systems \cite{Das2008} and consistently observed in chemically doped samples. In our boron-doped graphene system, the observed negative $(1/q)$ values further support the p-type nature of boron incorporation and confirm the presence of significant hole doping.
Recently symmetry breaking leading to structural instability in the charge density wave regime is reported for $2H-TaS_2$ \cite{Rawat2024}. From our estimates we find that the ((1)⁄(q)) factor changes from -0.25 to -0.27, thus implying a minute reduction of the EPC coupling right at temperatures $T_1$,$T_2$. We believe that this subtle change in the coupling strength and thereby triggering a fluctuation in the EPC coupling parameter can certainly lead to interesting charge ordering phenomena for our sample which is also observed for other Kagome lattice compounds \cite{He2024}. 

\section{Conclusion}
In conclusion, we have demonstrated that out-of-plane boron-doped reduced graphene oxide exhibits a quantum mechanical collective phenomenon characterized by a periodic distribution of correlated charges within the lattice. This lattice reconstruction near the transition temperature primarily arises from enhanced electron–phonon coupling (EPC) and the interaction between interlayer states and the $\pi$ electron density. Low-temperature Raman spectroscopy provides strong experimental support for the charge-ordered (CO) phase in the BG system. Temperature-dependent Raman analyses reveal subtle features in the G-band, including asymmetric line shapes well described by Breit–Wigner–Fano fitting, along with variations in peak position, FWHM, and EPC coupling strength $(1/q)$ near the transition temperatures—further validating the presence of charge ordering. Ab initio simulations corroborate these findings by revealing the associated electronic behavior that underpins the emergence of the ordered phase. Collectively, the spectroscopic and theoretical results offer a consistent explanation for the CO transition in this system.
 
\section{Supporting Information}
 Theoretical methods (Electronic Band structure), Details of BWF fit, Supplementary Tables, Supplementary Figures.
\section{Acknowledgments}
Author S. Sarkar thankfully acknowledges the Department of Science and Technology (DST), Govt. of India, for INSPIRE fellowship. Authors also thank University Grant Commission (UGC), Govt. of India, for the ‘University with potential for excellence-II’ scheme.  

\newpage
\hspace{1cm}
\hspace{1cm}
\begin{center}
\maketitle { \scalebox{1.2}{\textbf{Supporting Information}}  }
\end{center}

\widetext{
\section{Electronic Band Structure}
Supplementary figure S7(a) demonstrate the band dispersion of graphene containing different boron groups. For $BC_3$ and $BCO_2$ configurations, lowering of the associated Fermi levels from VBM are clearly seen, which signify the hole doping nature. For $BC_2O$ configuration, the Fermi level separates the VB and CB by opening a gap near itself. Another intriguing fact is that for $BCO_2$, the VB and CB resemble the shape of free electron configuration or $p_z$ electron as seen in pure graphene. In decomposed band structures of graphene with $BCO_2$ configuration, a large contribution from carbon $p_z$ orbital for construction of the VB is detected (Supplementary figure S7 (b)). Supplementary Figure S7 (c) represents the orbital decomposed band structure for the boron atoms in the formation of bands. For $BC_3$ groups, charge transfer from carbon to boron populates the $p_z$ orbitals and it becomes responsible for the formation of the bands near Fermi level. For $BC_2O$ groups, outermost $p_y$ electrons are involved to make up the VB, but for $BCO_2$ neither of these $p-$electrons are engaged in the band formation. Therefore, to achieve further evidence of the CO phase in graphene, more focused measurements are done for the BG samples which we demonstrate below.}

\section{Application of BWF Line Shape in Raman G Peak Analysis}

To accurately analyze the Raman G peak in boron-doped graphene, we employed the \textit{Breit--Wigner--Fano} (BWF) line shape fitting. The BWF model is particularly suited for capturing the asymmetric spectral profiles that arise due to Fano resonance. In pristine graphene, the G peak typically exhibits a symmetric Lorentzian profile, corresponding to the doubly degenerate in-plane vibrational $E_{2g}$ phonon mode at the Brillouin zone center. However, with the introduction of dopants such as boron, the electronic structure is significantly modified. These modifications result in the emergence of low-energy electronic continuum states that can interact with the G phonon mode, leading to asymmetry in the Raman peak. The BWF line shape effectively accounts for this interaction and is defined as\cite{Fano1866,Souvik2022}:

\begin{equation}
I(\omega) = I_0 + S \cdot \frac{\left[q + \dfrac{2(\omega - \omega_b)}{\Gamma}\right]^2}{1 + \left(\dfrac{2(\omega - \omega_b)}{\Gamma}\right)^2}
\end{equation}

Here, $\omega$ is the Raman shift, $\omega_b$ is the bare phonon frequency of the coupled mode, $\Gamma$ is the linewidth parameter related to the full-width at half maximum (FWHM), $q$ is the Fano asymmetry parameter (related to the ratio of resonant and non-resonant scattering amplitudes), $S$ is a scaling factor, and $I_0$ represents the background intensity. The parameter $|1/q|$ serves as a quantitative measure of electron-phonon coupling strength. Notably, in our boron-doped samples, we observe consistently negative values of $1/q$, indicative of hole doping, which aligns with the p-type nature of boron substitution in the graphene lattice. Furthermore, the magnitude of $1/q$ reflects the degree of interaction between the discrete phonon and the electronic continuum, providing deeper insights into the doping-induced modifications in the electronic and vibrational structure. Thus, the application of the BWF model enables a more accurate and physically meaningful interpretation of Raman spectral features in doped graphene systems, especially when conventional Lorentzian fitting fails to account for asymmetries arising from complex electron--phonon interactions.

\section{Tables and Figures}

\begingroup            
\begin{table*}[ht]
 \renewcommand\thetable{S1}
\caption {Binding energies and Full Width Half Maxima of all components found in B1s Peaks }
\begin{ruledtabular} 
\begin{tabular}{||c | c c c c c ||}
B.E. in eV (FWHM)/    & B-B       & $BC_3$     & $BC_2O$   & $BCO_2$  &  B-O \\
Sample Name &  in eV      & in eV    & in eV  & in eV  &  in eV  \\
\hline
BG500  &  -- (--)  & 188.8 (1.7)  &  190.2 (2.3) &  192.1 (2.5)  & 194.1 (2.4)  \\
BG700  &  186.8 (1.8)  & 188.7 (1.8)  &  190.1 (2.2) &  191.8 (2.4)  & 193.9 (2.4)  \\
BG900  &  186.9 (1.8)  & 188.5 (1.8)  &  189.7 (2.2) &  192.1 (2.4)  &  -- (--)  \\
BG1000 &  186.8 (1.9)  & 188.5 (1.9)  &  189.8 (2.1) &  191.7 (2.4)  & -- (--)  \\
               
\end{tabular}
\end{ruledtabular}
\end{table*}
\endgroup 

\begingroup            
\begin{center}
\begin{table*}[ht]
 \renewcommand\thetable{S2}
\caption {Amounts of different doping configurations of boron atoms}
\begin{ruledtabular} 
\begin{tabular}{||c | c c c |c ||}
Doping configuration/     & $BC_3$      & $BC_2O$   & $BCO_2$  &  Total out-of-plane boron  \\
Sample Name &  (\%)      & (\%)    & (\%)  & (\%)   \\
\hline

BG500  &  22 $\pm$ 3  & 63 $\pm$ 3  &  4 $\pm$ 1 &  89 $\pm$ 7   \\
BG700  &  43 $\pm$ 4  & 32 $\pm$ 4  &  4 $\pm$ 1 & 79 $\pm$ 9  \\
BG900  &  30 $\pm$ 4  & 9 $\pm$ 1  &  0 &  39 $\pm$ 5  \\
BG1000 &  23 $\pm$ 5  & 5 $\pm$ 2  &  0 &  27 $\pm$ 7  \\
               
\end{tabular}
\end{ruledtabular}
\end{table*}
\end{center}
\endgroup

\begingroup            	
\begin{center}
\begin{table*}
\renewcommand\thetable{S3}
\caption{Calculation of strain for differently configured BG samples.}
\begin{ruledtabular}

\begin{tabular}{|c|c|c|c|c|c|}
Doping & Bond type & Bond length & Pristine C-C & Strain $(\epsilon)$ in $(\%)$ & Max. Tensile \\
configuration & & in $\mathring{A}$ & in $\mathring{A}$ & & Strain in $(\%)$ \\ 
\hline
\multirow{6}{*}{BC$_{3}$} & B-C1(B) & 1.52  & 1.424 & 6.74157    & \multirow{6}{*}{10.95506} \\
                          & B-C2(B) & 1.58  & 1.424 & 10.95506   &  \\
                          & C1(B)-C & 1.381 & 1.424 & $-$3.01966 &  \\
                          & C1(B)-C & 1.4   & 1.424 & $-$1.68539 &  \\
                          & C2(B)-C & 1.423 & 1.424 & $-$0.07022 &  \\
                          & C2(B)-C & 1.49  & 1.424 & 4.63483    &  \\
\hline
\multirow{6}{*}{BC$_{2}$O} & B-C1(B) & 1.525 & 1.424 & 7.0927 &  \multirow{6}{*}{13.76404} \\
                           & B-C2(B) & 1.567 & 1.424 & 10.04213 &  \\
                          & C1(B)-C & 1.62 & 1.424 & 13.76404 &  \\
                          & C1(B)-C & 1.48 & 1.424 & 3.93258 &  \\
                          & C2(B)-C & 1.462& 1.424 & 2.66854 &  \\
                          & C2(B)-C & 1.528 & 1.424 & 7.30337 &  \\
\hline
\multirow{3}{*}{BCO$_{2}$} & B-C(B) & 1.639 & 1.424 & 15.09831 & 
\multirow{3}{*}{ 15.09831} \\
                          & C(B)-C & 1.491 & 1.424 & 4.70506  &  \\
                          & C(B)-C & 1.583 & 1.424 & 11.16573 &  \\
\end{tabular}
\end{ruledtabular}
\end{table*}
\end{center}
\endgroup 

\begingroup            	
\begin{center}
\begin{ruledtabular}
\begin{table}[ht]
\renewcommand\thetable{S4}
\centering
\caption{Temperature dependence of inverse asymmetry parameter ($1/q$), FWHM, and $\omega_G$ for G band in Raman spectra.}
\begin{tabular}{|c|c|c|c|}
\hline
\textbf{T (K)} & \textbf{$1/q$} & \textbf{FWHM (cm$^{-1}$)} & \textbf{$\omega_G$ (cm$^{-1}$)} \\
\hline
75  & -0.24761 $\pm$ 0.00564 & 63.26011 $\pm$ 0.8563  & 1610.5641 $\pm$ 0.39945 \\
\hline
80  & -0.25048 $\pm$ 0.00595 & 61.48249 $\pm$ 0.88809 & 1613.10956 $\pm$ 0.41323 \\
\hline
85  & -0.25638 $\pm$ 0.00549 & 60.57947 $\pm$ 0.80478 & 1613.73884 $\pm$ 0.37548 \\
\hline
90  & -0.25682 $\pm$ 0.00562 & 60.13616 $\pm$ 0.82023 & 1614.10965 $\pm$ 0.38266 \\
\hline
95  & -0.26663 $\pm$ 0.00589 & 61.22179 $\pm$ 0.85737 & 1613.9729 $\pm$ 0.40218 \\
\hline
100 & -0.25295 $\pm$ 0.00554 & 63.57801 $\pm$ 0.8458  & 1613.06352 $\pm$ 0.39317 \\
\hline
105 & -0.24669 $\pm$ 0.00574 & 63.36523 $\pm$ 0.88119 & 1612.7834 $\pm$ 0.40861 \\
\hline
110 & -0.26154 $\pm$ 0.00601 & 62.75871 $\pm$ 0.89828 & 1613.04613 $\pm$ 0.41991 \\
\hline
120 & -0.24801 $\pm$ 0.00599 & 63.9644 $\pm$ 0.92169  & 1612.11759 $\pm$ 0.42809 \\
\hline
130 & -0.26005 $\pm$ 0.00606 & 63.73793 $\pm$ 0.91712 & 1612.69125 $\pm$ 0.42802 \\
\hline
140 & -0.27235 $\pm$ 0.00589 & 63.32522 $\pm$ 0.87485 & 1613.6978 $\pm$ 0.41049 \\
\hline
\end{tabular}
\label{tab:raman_data}
\end{table}
\end{ruledtabular}
\end{center}
\endgroup

\begin{figure*}
\renewcommand\thefigure{S1}
\includegraphics[clip,width= 8.5 cm]{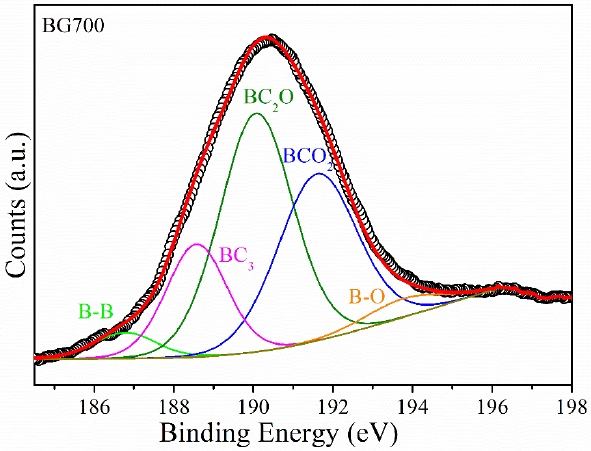}
\caption{Deconvolution of B1s spectra of BG700 sample into different configurations.}
\end{figure*}

\begin{figure*}
\renewcommand\thefigure{S2}
\includegraphics[clip,width= 8.5 cm]{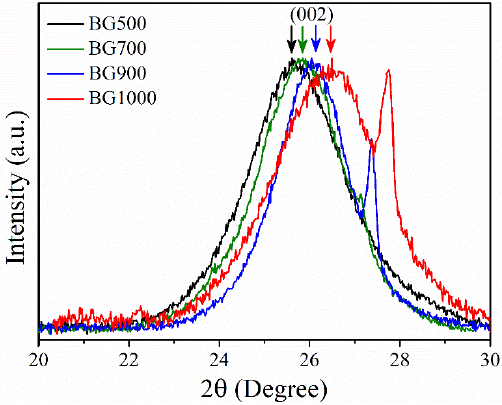}
\caption{XRD peaks for (002) plane of all BG samples.}
\end{figure*}

\begin{figure*}
\renewcommand\thefigure{S3}
\includegraphics[clip,width= 17 cm]{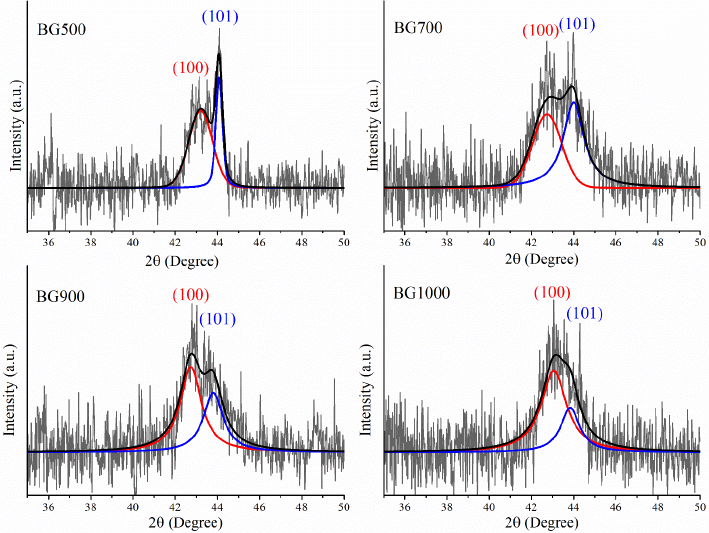}
\caption{XRD peaks for (100) and (101) planes of all BG samples}
\end{figure*}

\begin{figure*}
\renewcommand\thefigure{S4}
\includegraphics[clip,width= 17 cm]{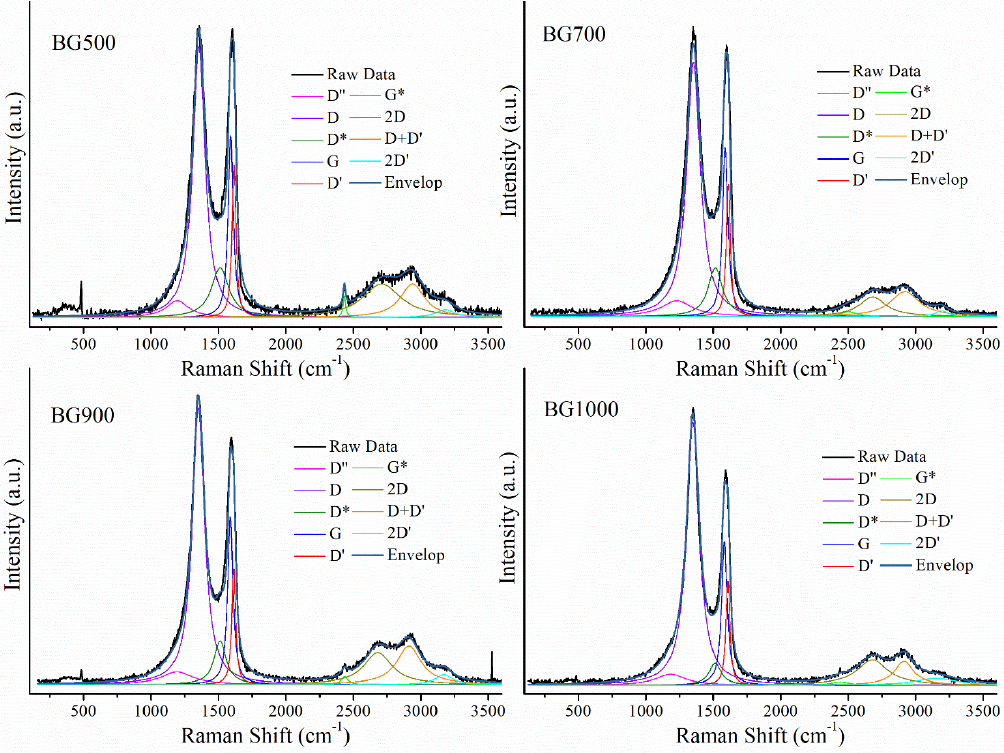}
\caption{Deconvoluted Raman spectra for all BG samples.}
\end{figure*}

\begin{figure*}
\renewcommand\thefigure{S5}
\includegraphics[clip,width= 6 cm]{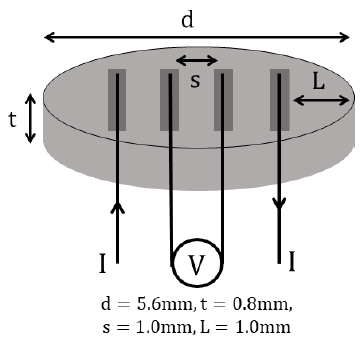}
\caption{Schematic for low temperature resistance measurement.}
\end{figure*}

\begin{figure*}
\renewcommand\thefigure{S6}
\includegraphics[clip,width= 17 cm]{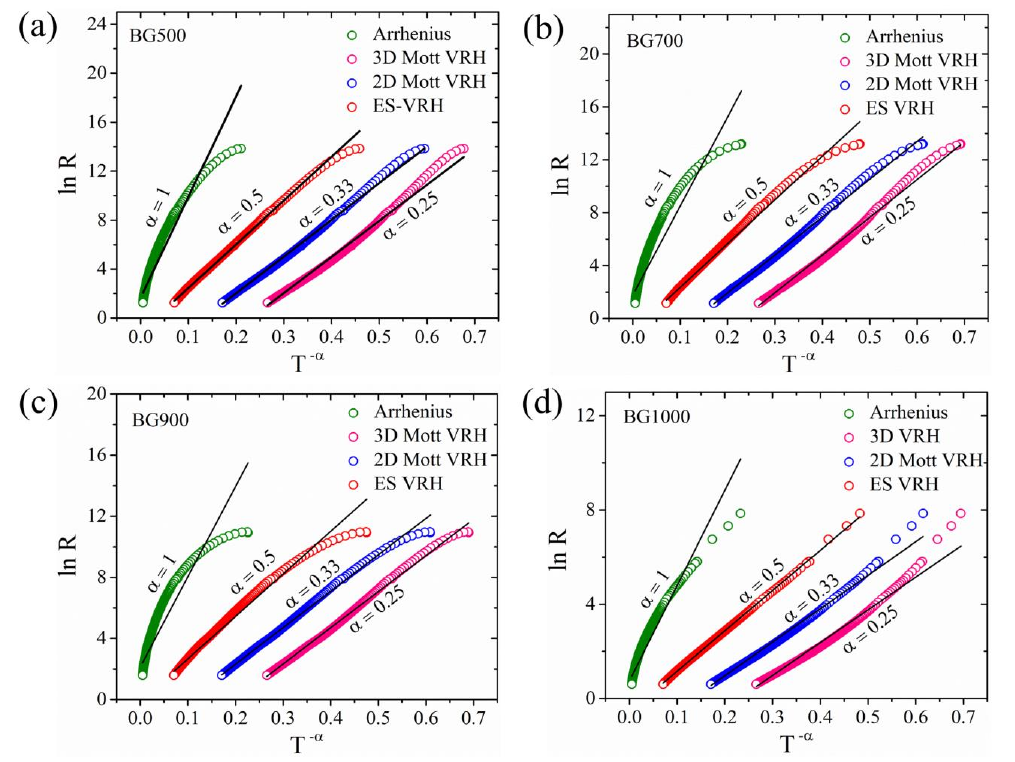}
\caption{Fit for possible mechanisms of carrier transport in BG samples as obtained from electrical R-T measurements.}
\end{figure*}

\begin{figure}
\renewcommand\thefigure{S7}
\includegraphics[clip,width= 15 cm]{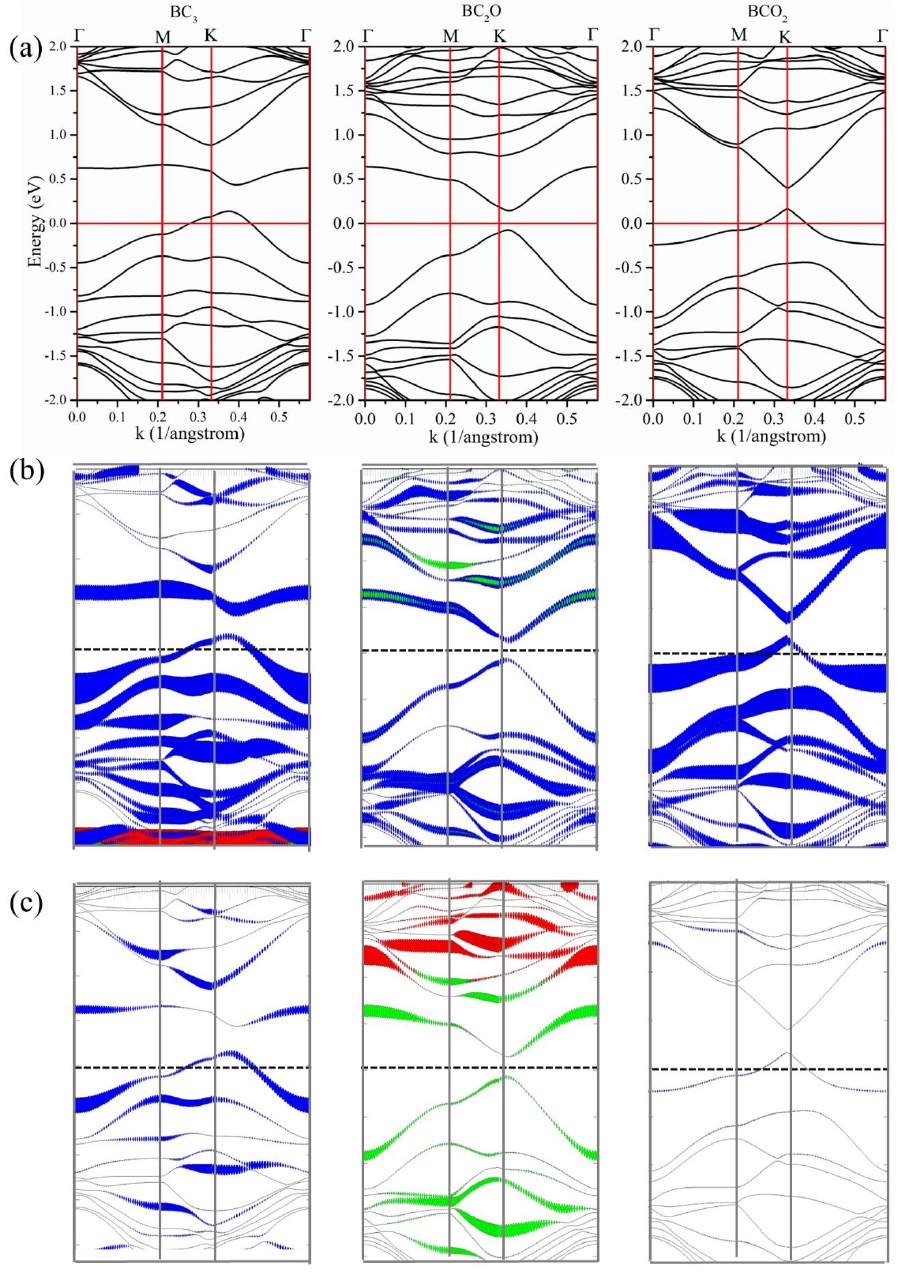}
\caption{Electronic band dispersion: (a) Band structure of graphene consists with $BC_3$ (left), $BC_2O$ (middle) and $BCO_2$ (right) configurations. (b) Decomposed band structure for Carbon atoms showing individual contributions of $p_x$(red), $p_y$(green), $p_z$(blue) orbitals for formation of bands. Thickness of the band lines indicates weightage of the contributions. (c) Decomposed band structure for boron atoms showing individual contributions of $p_x$(red), $p_y$(green), $p_z$(blue) orbitals for formation of bands. Thickness of the band lines indicates weightage of the contributions.}
\end{figure}

\FloatBarrier

\bibliography{reference_ms}
\bibliographystyle{apsrev4-2}


%

\end{document}